\documentclass[aps,reprint,superscriptaddress,twocolumn]{revtex4}

\usepackage{amsmath}
\usepackage{amssymb}
\usepackage{mathrsfs}
\usepackage{graphics}
\usepackage{microtype}

\usepackage[usenames,dvipsnames]{color}

\newcommand{\GF}{G_{\rm F}}
\newcommand{\rS}{{\rm S}}
\newcommand{\C}{\mathscr{C}}
\newcommand{\Sc}{\mathscr{S}}

\newcommand{\CC}{\mathcal{C}}

\newcommand{\sMA}[2]{\langle  #1,#2 \rangle}

\begin{document}


\title{\boldmath Standard-model prediction of $\epsilon_K$ with
  manifest CKM unitarity }

\author{Joachim~Brod}
\email[]{Joachim.Brod@uc.edu}
\affiliation{Department of Physics, University of Cincinnati, Cincinnati, OH 45221, USA}

\author{Martin~Gorbahn}
\email[]{Martin.Gorbahn@liverpool.ac.uk}
\affiliation{Department of Mathematical Sciences, University of Liverpool, Liverpool, L69 7ZL, UK}

\author{Emmanuel~Stamou}
\email[]{Emmanuel.Stamou@epfl.ch}
\affiliation{Institut de Th\'eorie des Ph\'enomenes Physiques, EPFL, Lausanne, Switzerland}

\date{\today}

\begin{abstract}
The parameter $\epsilon_K$ describes CP violation in the neutral kaon
system and is one of the most sensitive probes of new physics. The
large uncertainties related to the charm-quark contribution to
$\epsilon_K$ have so far prevented a reliable standard-model
prediction. We show that CKM unitarity enforces a unique form of the
$|\Delta\rS\!=\!2|$ weak effective Lagrangian in which the
short-distance theory uncertainty of the imaginary part is
dramatically reduced. The uncertainty related to the charm-quark
contribution is now at the percent level.  We present the updated
standard-model prediction $\epsilon_K = 2.16(6)(8)(15)\times10^{-3}$,
where the errors in brackets correspond to QCD short-distance and
long-distance, and parametric uncertainties, respectively.

\end{abstract}


\maketitle

\section{Introduction}

CP violation in the neutral kaon system, parameterized by
$\epsilon_K$, is one of the most sensitive precision probes of new
physics. For decades, the large perturbative uncertainties related to
the charm-quark contributions have been an impediment to fully
exploiting the potential of $\epsilon_K$. In this letter we
demonstrate how to overcome this obstacle.

The parameter $\epsilon_K$ can be defined as~\cite{Anikeev:2001rk}
\begin{equation}\label{eq:ek:def}
  \epsilon_K \equiv e^{i\phi_\epsilon} \sin\phi_\epsilon \frac{1}{2}
  \arg \bigg( \frac{-M_{12}}{\Gamma_{12}} \bigg)\,.
\end{equation}
Here, $\phi_\epsilon = \arctan(2\Delta M_K/\Delta\Gamma_K)$, with
$M_K$ and $\Delta\Gamma_K$ the mass and lifetime difference of the
weak eigenstates $K_L$ and $K_S$. $M_{12}$ and $\Gamma_{12}$ are the
Hermitian and anti-Hermitian parts of the $|\Delta\rS\!=\!2|$ weak
effective Hamiltonian. The short-distance contributions to
$\epsilon_K$ are then contained in the matrix element $M_{12} \equiv
\langle K^0 | \mathcal{H}^{\Delta\rS = 2}_{f=3}| \bar K^0 \rangle /
(2\Delta M_K)$. Both $M_{12}$ and $\Gamma_{12}$ depend on the phase
convention of the Cabibbo--Kobayashi--Maskawa (CKM) matrix $V$.
To make the cancellation of the phase
convention in Eq.~\eqref{eq:ek:def} explicit, we define the effective
$|\Delta\rS\!=\! 2|$ Hamiltonian in the three-quark theory as
\begin{equation}\label{eq:HS2:inv}
\begin{split}
	&\mathcal{H}_{f=3}^{\Delta\rS= 2} = \frac{\GF^2M_W^2}{4\pi^2} 
  \frac{1}{(\lambda_u^*)^2}  {Q}_{{\rS}2} 
  \Big\{ f_{1} \,\CC_{1}(\mu) \\
   & + i J \left[ f_{2} \,\CC_{2}(\mu) 
                + f_{3}\,\CC_{3}(\mu) \right] \Big\} 
     + \text{h.c.} + \ldots
\end{split}
\end{equation}
in terms of the real Wilson coefficients $\CC_{i}(\mu)$, $i=1,2,3$,
and four real, independent, rephasing-invariant parameters $J$,
$f_{1}$, $f_{2}$, and $f_{3}$ comprising the CKM matrix
elements. Here, we defined $\lambda_i \equiv V_{is}^* V_{id}$. The
local four-quark operator
\begin{equation}\label{eq:def:QS2} Q_{{\rS}2} = 
	(\overline{s}_L \gamma_{\mu} d_L) \otimes
(\overline{s}_L \gamma^{\mu}d_L)\,,
\end{equation}
defined in terms of the left-handed $s$- and $d$-quark fields, induces
the $|\Delta\rS\!=\!2|$ transitions. The ellipsis in
Eq.~\eqref{eq:HS2:inv} represents $|\Delta\rS\!=\!1|$ operators that
contribute to the dispersive and absorptive parts of the amplitude via
non-local insertions, as well as operators of mass dimension higher
than six.

The normalization factor $1/(\lambda_u^*)^2$ in Eq.~\eqref{eq:HS2:inv}
ensures that the resulting expression of $\epsilon_K$ in
Eq.~\eqref{eq:ek:def} is phase-convention independent if one
accordingly extracts the factor $1/\lambda_u^*$ from the
$|\Delta\rS\!=\!1|$ Hamiltonian which contributes to $\Gamma_{12}$ via
a double insertion.  Moreover, the splitting into the real and
imaginary part in Eq.~\eqref{eq:HS2:inv} is unique. Explicitly, we
have $J = \mathrm{Im}(V_{us} V_{cb} V^*_{ub} V^*_{cs})$ and $f_{1} =
|\lambda_u|^4 + \dots$, where the ellipsis denotes real terms that are
suppressed by powers of the Wolfenstein parameter $\lambda$.

By contrast, the splitting of the imaginary part among
$f_2$ and $f_3$ is not
unique. A particularly convenient choice is $f_{2} = 2
\mathrm{Re}(\lambda_t \lambda_u^*)$ and $f_{3} = |\lambda_u|^2$,
leading to the Lagrangian
\begin{equation}\label{eq:LS2:final}
\begin{split}
	\mathcal{L}^{\Delta\rS=2}_{f=3} = - \frac{\GF^2 M_W^2}{4 \pi^2}
    \big[ & \lambda_u^2 \C_{{\rS}2}^{uu}(\mu) + \lambda_t^2 \C_{{\rS}2}^{tt}
    (\mu) \\ +& \lambda_u \lambda_t \C_{{\rS}2}^{ut}(\mu) \big]  Q_{{\rS}2}
      + \textrm{h.c.} + \dots \,,
\end{split}
\end{equation}
where we used CKM unitarity and identified $\C_{{\rS}2}^{uu} \equiv
\CC_1$, $\C_{{\rS}2}^{tt} \equiv \CC_2$, and $\C_{{\rS}2}^{ut} \equiv
\CC_3$.  This form of the effective Lagrangian, where the coefficient
of $\C_{{\rS}2}^{uu}$ is real, has been suggested in
Ref.~\cite{Christ:2012se} as a better way to compute the matrix
elements on the lattice in the four-flavor theory, and it was
speculated that also the perturbative part may then converge better.
Above, we showed that this minimal form is essentially dictated by CKM
unitarity; we will see below that, indeed, both $\CC_2$ and $\CC_3$
(as opposed to $\CC_1$!) have a perfectly convergent perturbative
expansion.

Traditionally, however, the effective Lagrangian has been 
given in a different form~\cite{Buchalla:1995vs, Herrlich:1996vf},
\begin{equation}\label{eq:LS2:conv:useful}
\begin{split}
	\mathcal{L}^{\Delta\rS=2}_{f=3} = - \frac{\GF^2 M_W^2}{4 \pi^2} \big[
      & \lambda_c^2 C_{{\rS}2}^{cc}(\mu) +
      \lambda_t^2 C_{{\rS}2}^{tt}(\mu)\\ 
      +&
      \lambda_c \lambda_t C_{{\rS}2}^{ct}(\mu) \big]
       Q_{{\rS}2} + \textrm{h.c.} + \dots \,,
\end{split}
\end{equation}
which can be obtained from Eq.~\eqref{eq:HS2:inv} via the choice $f_2
= \mathrm{Re}(\lambda_t \lambda_u^*)$, $f_3 = \mathrm{Re}(\lambda_c
\lambda_u^*)$, where we are now lead to identify $C_{{\rS}2}^{cc}
\equiv \CC_1$, $C_{{\rS}2}^{ct} \equiv 2\CC_1 + \CC_3$, and
$2C_{{\rS}2}^{tt} \equiv 2\CC_1 + \CC_2 + \CC_3$. We see that in this
choice $\CC_1$ artificially enters all three coefficients, which all
contribute to $\epsilon_K$. This is unfortunate because the
perturbative expansion of $\CC_1$ exhibits bad convergence, as shown
in Ref.~\cite{Brod:2011ty}.

Clearly, Eq.~\eqref{eq:LS2:final} can be directly obtained from
Eq.~\eqref{eq:HS2:inv} by the replacement $\lambda_u = -\lambda_c
-\lambda_t$. We will refer to Eq.~\eqref{eq:LS2:conv:useful} as
``$c$-$t$ unitarity'' and to Eq.~\eqref{eq:LS2:final} as ``$u$-$t$
unitarity''. It is customary to define the
renormalization-scale-invariant (RI) Wilson coefficients
$\widehat{C}_{{\rS}2}^{ij} \equiv C_{{\rS}2}^{ij}(\mu) b(\mu)$, $ij =
cc, ct, tt$, where the scale factor $b(\mu)$ is defined, for instance,
in Refs.~\cite{Herrlich:1996vf, Brod:2010mj}.  QCD corrections are
then parameterized by the factors $\eta_{tt}$, $\eta_{ct}$, and
$\eta_{cc}$, defined in terms of the Inami--Lim functions $S(x_i,
x_j)$ (see Ref.~\cite{Inami:1980fz}) by $\widehat{C}_{{\rS}2}^{tt}
=\eta_{tt} S(x_t)$, $\widehat{C}_{{\rS}2}^{ct} = 2 \eta_{ct}
S(x_c,x_t)$, and $\widehat{C}_{{\rS}2}^{cc} = \eta_{cc} S(x_c)$. Here,
we defined the mass ratios $x_i \equiv m_i(m_i)^2/M_W^2$ with
$m_i(m_i)$ denoting the RI $\overline{\rm MS}$ mass.  $\eta_{tt}$ is
known at next-to-leading-logarithmic (NLL) order in QCD, $\eta_{tt} =
0.5765(65)$~\cite{Buras:1990fn}, while the other two are known at
next-to-next-to-leading-logarithmic (NNLL) order, $\eta_{ct} =
0.496(47)$~\cite{Brod:2010mj} and $\eta_{cc} =
1.87(76)$~\cite{Brod:2011ty}.

In the same way, we define the RI Wilson coefficients and the QCD
correction factors for the Lagrangian in Eq.~\eqref{eq:LS2:final},
namely, $\widehat{\C}_{{\rS}2}^{tt} = \eta_{tt} \Sc(x_t)$ and
$\widehat{\C}_{{\rS}2}^{ut} = 2 \eta_{ut} \Sc(x_c,x_t)$. Note that
since $\C_{{\rS}2}^{uu}$ is real, it is not required to obtain
$\epsilon_K$. Using Eqs.~\eqref{eq:LS2:final}
and~\eqref{eq:LS2:conv:useful} and the unitarity relation $\lambda_c =
-\lambda_u -\lambda_t$, it is readily seen that the modified
Inami--Lim functions $\Sc(x_i, x_j)$ are given by $\Sc(x_c) = S(x_c)$,
$\Sc(x_c,x_t) = S(x_c) - S(x_c,x_t)$, and $\Sc(x_t) = S(x_t) + S(x_c)
- 2S(x_c,x_t)$. The latter relation implies that $\eta_{tt}$ coincides
in $u$-$t$ and $c$-$t$ unitarity up to tiny corrections of order
${\cal O}(m_c^2/M_W^2) \sim 10^{-4}$, which we neglect. In what
follows, we show that $\eta_{ut} = 0.402(5)$ at NNLL, with an
order-of-magnitude smaller uncertainty than $\eta_{ct}$ and
$\eta_{cc}$.

\section{Analytic results\label{sec:weakeffective}}

In this section we will show that all ingredients for the NNLL analysis
with manifest CKM unitarity of the charm contribution to $\epsilon_K$
are available in the literature. To establish the requisite relations,
we display the effective five- and four-flavor Lagrangian using both
the traditional $c$-$t$ unitarity, giving~\cite{Herrlich:1996vf, Brod:2010mj}
\begin{align}
	\label{eq:lag:s2:ct}
	&\mathcal{L}_{f=4,5}^{\text{eff}} =\nonumber\\
	&- \frac{4 \GF}{\sqrt{2}} \biggl(
		\sum_{k,l=u,c} \!\! V_{ks}^\ast V_{ld} (C_{+} Q_+^{kl} + C_{-} Q_-^{kl})
                - \lambda_t \sum_{i=3,6} C_i Q_i\biggr)\nonumber\\
	&- \frac{\GF^2 M_W^2}{4\pi^2} \lambda_t^2 {C}_{\text{{\rS}2}} {Q}_{\text{{\rS}2}}
	- 8 \GF^2 \lambda_c \lambda_t \tilde{C}_7 \tilde{Q}_{7} + \text{h.c.}\, .
\end{align}
and $u$-$t$ unitarity, giving
\begin{align}
	\label{eq:lag:s2:ut}
	&\mathcal{L}_{f=4,5}^{\text{eff}} =\nonumber\\
	&- \frac{4 \GF}{\sqrt{2}} \biggl(
		\sum_{k,l=u,c} \!\!\! V_{ks}^\ast V_{ld} (\C_{+} Q_+^{kl} + \C_{-} Q_-^{kl})
                 - \lambda_t \sum_{i=3,6} \C_i Q_i\biggr)\nonumber\\
	&- \frac{\GF^2 M_W^2}{4\pi^2} \lambda_t^2 {\C}_{\text{{\rS}2}} {Q}_{\text{{\rS}2}}
	- 8 \GF^2 (\lambda_u \lambda_t +\lambda_t^2  )\tilde{\C}_7  \tilde{Q}_{7} 
         + \text{h.c.}
\end{align}
The Wilson coefficients in Eqs.~\eqref{eq:lag:s2:ut}
and~\eqref{eq:lag:s2:ct} are related via
\begin{align}\label{eq:c:relation}
	&\C_{i} = C_{i}\,,&
	&{\C}_{\text{{\rS}2}} = {C}_{\text{{\rS}2}}\,,&
	&\tilde{\C}_{7} = -\tilde{C}_{7}\,,&
\end{align}
where $i=+,-,3,\dots,6$. Here, $\tilde{Q}_{7} \equiv m_c^2/g_s^2
{Q}_{{\rS}2}$, with $g_s$ the strong coupling constant, while the
remaining operators (current--current and penguin operators) are
defined in Ref.~\cite{Brod:2010mj}. The initial conditions for all the
$C_i$ Wilson coefficients and $\tilde{C}_{7}$, up to NNLO, can be
found in Refs.~\cite{Buras:1991jm, Bobeth:1999mk, Buras:2006gb,
  Brod:2010mj}.

It is evident that the renormalization-group evolution of the 
coefficients $\C_{i}$ and
$C_{i}$, as well as of ${\C}_{\text{{\rS}2}}$ and
${C}_{\text{{\rS}2}}$, is identical. We now show that also the
mixing of the $\C_{i}$ into $\tilde{C}_{7}$ via double insertions of
dimension-six operators can be obtained from results available in the
literature. To this end we define the following short-hand notation
for the relevant $|\Delta\rS\!=\!2|$ matrix elements of double insertions
of local operators $O_A$ and $O_B$,
\begin{equation}
\langle O_A, O_B \rangle \equiv \frac{i^2}{2!} \int d^4x d^4y \langle
T\{O_A(x) O_B(y)\}\rangle \,.
\end{equation}

With the Lagrangian in Eq.~\eqref{eq:lag:s2:ct} and using
$(V_{cs}^*V_{ud})(V_{us}^*V_{cd}) = -\lambda_c^2 - \lambda_c\lambda_t$,
the anomalous dimensions for the mixing of two $\C_{i}$s into
$\tilde{C}_{7}$ can then be obtained from the divergent part of the
amplitude
\begin{align}\label{eq:double}
		&{\cal M}_\text{double insertions}^{\Delta\rS=2}\big|_\text{div} \\[0.5em]
	\propto& ~ ~\lambda_t^2 \left( \sMA{Q_P}{Q_P}
		           +\sMA{Q^{uu}}{Q^{uu}}
			   +2\sMA{Q_{P}}{Q^{uu}}
	              \right)\big|_\text{div}\nonumber\\
	-& \lambda_c\lambda_t \big( 
			     2 \sMA{Q_P}{Q^{cc} -Q^{uu} }
			     + \sMA{Q^{cc}}{Q^{cc}}
        - \sMA{Q^{uu}}{Q^{uu}}\big)\big|_\text{div}\nonumber\\[0.5em]
	   =& ~~\lambda_t^2 \left( \sMA{Q_P}{Q_P}
		           +\sMA{Q^{cc}}{Q^{cc}}
			   +2\sMA{Q_{P}}{Q^{cc}}
	              \right)\big|_\text{div}\nonumber\\
	+&\lambda_u\lambda_t \big( 
			     2 \sMA{Q_P}{Q^{cc} -Q^{uu} }
			     + \sMA{Q^{cc}}{Q^{cc}}
        - \sMA{Q^{uu}}{Q^{uu}}
\big)\big|_\text{div}.\nonumber
\end{align}
We introduced the short-hand notations $Q_P \equiv \sum_{i=3}^6
\C_iQ_i$ and $Q^{qq'} \equiv \sum_{i=+,-} \C_iQ_i^{qq'}$. In the first
equality we utilized the observation that the
divergence of the linear combination of amplitudes proportional to
$\lambda_c^2$ vanishes~\cite{Witten:1976kx},
\begin{equation}\label{eq:witten}
\left( \sMA{Q^{cc}-Q^{uu}}{Q^{cc} -Q^{uu} } - 2\sMA{Q^{uc}}{Q^{cu}}
\right)\big|_\text{div} = 0\,.
\end{equation} 
In the second equality we used, in addition, the unitarity relation
$\lambda_c = -\lambda_u -\lambda_t$. We see that the divergent parts
of the amplitudes proportional to $\lambda_c\lambda_t$ and
$\lambda_u\lambda_t$ are the same up to a sign. Therefore, the
corresponding anomalous dimensions can be extracted from existing
literature. In the notation of Ref.~\cite{Brod:2010mj} we have $\tilde
\gamma_{\pm,7}^{(ut)} = - \tilde \gamma_{\pm,7}^{(ct)}$, where the
superscripts ``$ut$'' and ``$ct$'' denote the results in $u$-$t$
unitarity and $c$-$t$ unitarity, respectively. All other contributing
anomalous dimensions remain unchanged.

Note that in the second equality in Eq.~\eqref{eq:double}, the
amplitudes proportional to $\lambda_t^2$ involve the charm-flavored
current-current operators. This is related to the appearance of an
initial condition of the operator $\tilde{Q}_7$ at the weak scale
proportional to $\lambda_t^2$. This charm-quark contribution to
${\C}_{{\rS}2}^{tt}$ will be neglected in this work, as
discussed above. In this approximation, ${\C}_{{\rS}2}^{tt}$
is identical to ${C}_{{\rS}2}^{tt}$ and can be directly taken
from the literature~\cite{Buras:1990fn}.

Also the matching of the four- onto the three-flavor effective
Lagrangian at $\mu_c$ changes in a simple way. Picking the coefficient of
$\lambda_u\lambda_t$, the matching of the Lagrangian in 
Eq.~\eqref{eq:lag:s2:ut} onto the one in 
Eq.~\eqref{eq:LS2:final} yields the condition
\begin{align}
\sum_{i,j=+,-} & \C_i(\mu_c) \C_j(\mu_c) \Big(2\langle Q_{i}^{cc}, Q_j^{cc} \rangle \nonumber \\ 
  & -2\langle Q_{i}^{uc},
  Q_j^{cu} \rangle -2\langle Q_{i}^{uu}, Q_j^{cc} \rangle \Big)(\mu_c) \label{eq:match:c:ut}\\ 
  + \sum_{i=3}^6 & \sum_{j=+,-}
  \C_i(\mu_c) \C_j(\mu_c) 2\langle Q_{i}, Q_j^{cc} - Q_j^{uu} \rangle(\mu_c) \nonumber\\
  & + \tilde{\C}_{7}(\mu_c) \langle {\tilde Q}_7
  \rangle(\mu_c) = \frac{1}{32\pi^2} {\C}_{{\rS}2}^{
    ut}(\mu_c) \langle { Q}_{{\rS}2} \rangle(\mu_c) \,.\nonumber
\end{align}
Alternatively, selecting the coefficient of $\lambda_c\lambda_t$, the
matching of the Lagrangian in Eq.~\eqref{eq:lag:s2:ct} onto the one in
Eq.~\eqref{eq:LS2:conv:useful} yields the condition
\begin{align}
 \sum_{i,j=+,-} & C_i(\mu_c) C_j(\mu_c) \Big( 2\langle Q_{i}^{uu}, Q_j^{uu} \rangle  \nonumber\\ 
	 & -2\langle Q_{i}^{uc}, Q_j^{cu} \rangle -2\langle Q_{i}^{uu}, Q_j^{cc} \rangle \Big)(\mu_c)\label{eq:match:c:ct} \\
  + \sum_{i=3}^6 & \sum_{j=+,-}
  C_i(\mu_c) C_j(\mu_c) 2\langle Q_{i}, Q_j^{uu} - Q_j^{cc} \rangle(\mu_c) \nonumber\\
  & + {\tilde C}_7(\mu_c) \langle {\tilde Q}_7
  \rangle(\mu_c) = \frac{1}{32\pi^2} { C}_{{\rS}2}^{
    ct}(\mu_c) \langle { Q}_{{\rS}2} \rangle(\mu_c) \,.\nonumber
\end{align}
and for the coefficient of $\lambda_c^2$ yields the condition
\begin{equation}\label{eq:match:c:cc}
\begin{split}
  & \sum_{i,j=+,-} C_i(\mu_c) C_j(\mu_c) \big(\langle Q_{i}^{cc} - Q_{i}^{uu},
  Q_j^{cc} - Q_j^{uu} \rangle \\ & -2\langle Q_{i}^{uc}, Q_j^{cu} \rangle \big)(\mu_c) 
  = \frac{1}{32\pi^2} {C}_{{\rS}2}^{cc}(\mu_c) \langle { Q}_{{\rS}2} \rangle(\mu_c) \,.
\end{split}
\end{equation}
Recalling Eq.~\eqref{eq:c:relation}, we see that 
${\C}_{{\rS}2}^{ut} = 2{C}_{{\rS}2}^{cc} - {C}_{{\rS}2}^{ct}$, hence
we can extract also the matching conditions from the literature.

In order to provide the explicit expressions, we parameterise the
operator matrix elements as:
\begin{equation}
\begin{split}
  \langle {\tilde Q}_7 \rangle = r_{7} \langle {\tilde Q}_{7}
  \rangle^{(0)}\, , \quad \langle { Q}_{{\rS}2} \rangle = r_{{\rS}2} \langle
  { Q}_{{\rS}2} \rangle^{(0)} \,,\\ \quad \langle
  Q_i Q_j \rangle^{qq'} (\mu_c) = \frac{1}{32\pi^2}\frac{m_c^2(\mu_c)}{M_W^2}
  r_{ij,\rS 2}^{qq'} \langle { Q}_{{\rS}2} \rangle^{(0)}  \,. 
\end{split}
\end{equation}
Here, the superscripts $qq' = ut, ct, cc$ denote the specific
flavor structures appearing in the double insertions in
Eqs.~\eqref{eq:match:c:ut},~\eqref{eq:match:c:ct},
and~\eqref{eq:match:c:cc}, respectively. 
The matching contributions
are then given in terms of the literature results by 
$r_{ij,\rS 2}^{ut} =
2r_{ij,\rS 2}^{cc} - r_{ij,\rS 2}^{ct}$. 
It is interesting to note that, due
to the presence of a large logarithm $\log(m_c/M_W)$ in the function
$\Sc(x_c, x_t)$, only the NLO result for $\eta_{cc}$ of
Ref.~\cite{Herrlich:1993yv} is required. The remaining NNLO results
can be found in Refs.~\cite{Herrlich:1996vf, Brod:2010mj}.

\section{Numerics}

\begin{figure*}[]
	\centering
	\includegraphics{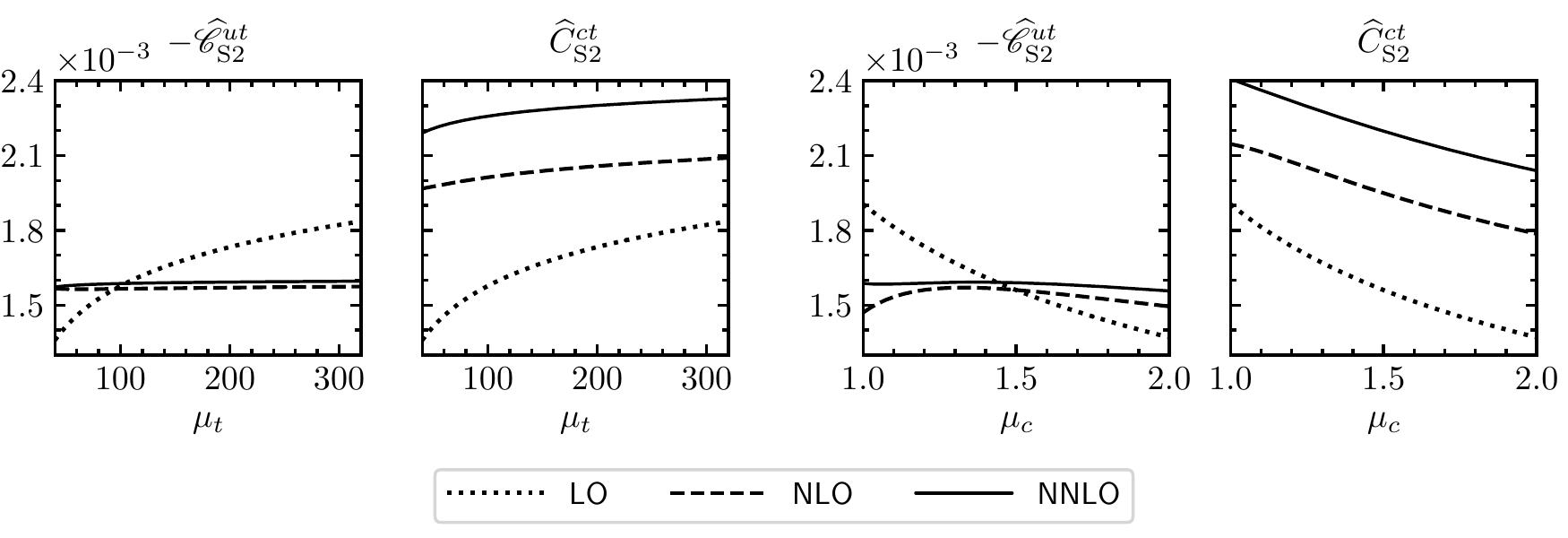}\\[-0.5em]
	\caption{
		Comparison of Wilson coefficients in $u$-$t$ ($1^{\text{st}}$ and $3^{\text{rd}}$ plot)
	and $c$-$t$ unitarity ($2^{\text{nd}}$ and $4^{\text{th}}$ plot).
	Shown is the residual
          renormalization-scale dependence of the RI Wilson
	  coefficients as a proxy for their theory uncertainty.
	  In the two plots on the left the five-flavour threshold, $\mu_t$, 
	  is varied, while in the two on the right the three-flavour threshold, $\mu_c$,
	  is varied (see text for further details).
\label{fig:scalevariation}}
\end{figure*}

In Sec.~\ref{sec:weakeffective} we extracted all the necessary
quantities to evaluate the $\lambda_t^2$ and $\lambda_u\lambda_t$
contributions to $\epsilon_K$ at NLL and NNLL accuracy, respectively.
Here, we discuss the residual theory uncertainties in $u$-$t$
unitarity and compare them to the traditional approach of $c$-$t$
unitarity.  To estimate the uncertainty from missing, higher-order
perturbative corrections we vary the unphysical thresholds $\mu_t$,
$\mu_b$, and $\mu_c$ in the ranges $40\,\text{GeV} \!\leq \!\mu_t
\!\leq\! 320\,\text{GeV}$, $2.5\,\text{GeV} \!\leq \!\mu_b \!\leq\!
10\,\text{GeV}$, and $1\,\text{GeV} \!\leq \!\mu_t \!\leq\!
2\,\text{GeV}$.  When varying one scale we keep the other two scales
fixed at the values of the RI mass of the fermions, $\mu_i=m_i(m_i)$
with $i=t,b,c$.  The central values for the $\eta$ parameters are
obtained as the average between the lowest and highest value of the
three scale variations, and their scale uncertainty as half the
difference of the two values.  The leading, but small, parametric
uncertainties of $\alpha_s$ and $m_c$ are obtained by varying the
parameters at their respective $1\sigma$ ranges.  We find
\begin{align}
	\label{eq:etanum}
	\eta_{tt}^{\rm NLL} &= \phantom{0}0.55 (1\pm4.2\%_{\text{scales}} \pm 0.1\%_{\alpha_s})\,,\\
	\eta_{ut}^{\rm NNLL}&= 0.402(1\pm1.3\%_{\text{scales}} \pm0.2\%_{\alpha_s}\pm0.2\%_{m_c})\,.\nonumber
\end{align}

Apart from the tiny correction of ${\cal O}(m_c^2/M_W^2) \sim 10^{-4}$
$\eta_{tt}$ is not affected by the different choice of CKM
unitarity. The difference in the scale uncertainty with respect to
Ref.~\cite{Buras:1990fn} is mainly due to the larger range of scale
variation chosen here.  By contrast, the residual scale uncertainty of
$\eta_{ut}$ is significantly less than the corresponding one in
$\eta_{ct}$ and $\eta_{ct}$ in $c$-$t$ unitarity. To illustrate this,
we show in Fig.~\ref{fig:scalevariation} the RI invariant Wilson
coefficients $\widehat{\C}^{ut}$ and $\widehat{C}^{ct}$ as a function
of the unphysical thresholds $\mu_t$ (left two panels) and $\mu_c$
(right two panels).

To obtain the standard-model prediction for $\epsilon_K$ we employ the
Wolfenstein parameterization~\cite{Tanabashi:2018oca} of the CKM factors in
Eq.~\eqref{eq:LS2:final}.
In the leading approximation we find $\text{Im}(\lambda_t^2) =
-2\lambda^{10}A^4\bar\eta(1-\bar\rho) + {\mathcal O}(\lambda^{12})$
and $\text{Im}(\lambda_u \lambda_t) = \lambda^6 A^2 \bar\eta +
{\mathcal O}(\lambda^{10})$. Numerically, the neglected terms amount
to sub-permil effects and can be safely neglected.  Therefore, we can
use the phenomenological expression (cf. Refs.~\cite{Buchalla:1995vs,
  Buras:2008nn, Buras:2010pza})
\begin{equation}\label{eq:eKformula}
\begin{split}
|\epsilon_K|
= & \kappa_\epsilon C_\epsilon \widehat{B}_K
|V_{cb}|^2 \lambda^2 \bar \eta \\ & \times \Big(|V_{cb}|^2(1-\bar\rho)
\eta_{tt} \Sc(x_t) - \eta_{ut} \Sc(x_c, x_t) \Big)\,,
\end{split}
\end{equation}
where
\begin{equation}\label{eq:Ce}
C_\epsilon = \frac{\GF^2 F_K^2 M_{K^0} M_W^2}{6\sqrt{2}\pi^2\Delta M_K}\,.
\end{equation}
We write $\bar\eta = R_t \sin\beta$ and $1-\bar\rho = R_t \cos\beta$,
with the quantity $R_t$ given by
\begin{equation}\label{eq:Rt}
R_t \approx \frac{\xi_s}{\lambda} \sqrt{\frac{M_{B_s}}{M_{B_d}}}
\sqrt{\frac{\Delta M_d}{\Delta M_s}}\,.
\end{equation}
Here, $\xi_s =
(F_{B_s}\sqrt{\widehat{B}_s})/(F_{B_d}\sqrt{\widehat{B}_d}) =
1.206(17)$ is a ratio of $B$-meson decay constants and bag factors
that is computed on the lattice~\cite{Aoki:2019cca}. The kaon bag
parameter is given by $\widehat{B}_K =
0.7625(97)$~\cite{Aoki:2019cca}. The phenomenological parameter
$\kappa_\epsilon = 0.94(2)$~\cite{Buras:2010pza} comprises
long-distance contributions not included in $B_K$.  As input for the
top-quark mass we use RI $\overline{\rm MS}$ mass $m_t(m_t) =
163.48(86)\,$GeV.  We obtain it by converting the pole mass $M_t =
173.1(9)\,$GeV~\cite{Tanabashi:2018oca} to $\overline{\rm MS}$ at
three-loop accuracy using {\tt RunDec}~\cite{Chetyrkin:2000yt}.  All
remaining numerical input is taken from Ref.~\cite{Tanabashi:2018oca}.

Using the $\eta$ values in Eq.~\eqref{eq:etanum} and adding 
errors in quadrature we find the standard-model
prediction
\begin{equation}
\begin{split}
  |\epsilon_K| & = \big( 
	  2.161 
		\pm 0.153_\text{param.} 
	   	\pm 0.064_{\eta_{tt}} 
	        \pm 0.008_{\eta_{ut}} 
		\\ & \qquad 
		\pm 0.027_{\widehat{B}_K} 
	        \pm 0.052_{\xi_s} 
	        \pm 0.046_{\kappa_\epsilon} \big)
		\times 10^{-3} \,,\\
               & = \big( 
		       2.161 
		\pm 0.153_\text{param.}\\
		&\qquad 
		\pm 0.076_\text{non-pert.}
		\pm 0.065_\text{pert.} 
		\big)\times 10^{-3} \,,\\
		& = 2.16(18)\times 10^{-3} \,.
\end{split}
\end{equation}
We see that the perturbative uncertainty ($\sim\!\!3.0\%$)
is now of the same order as the combined non-perturbative one ($\sim\!\!3.5\%$),
while the dominant uncertainties originate from the parametric, 
experimental uncertainties ($\sim\!\!7.1\%$).
Moreover, the dominant perturbative uncertainty no longer originates
from $\eta_{ct}$ but from the top-quark contribution, $\eta_{tt}$.

\section{Discussion and Conclusions}
In this letter, we showed that a manifest implementation of CKM
unitarity in the effective $|\Delta\rS\!=\! 2|$ Hamiltonian
dramatically improves the convergence behaviour of the perturbative
series for its imaginary part, by removing a spurious long-distance
charm-quark contribution.  In this way, and using only known results
in the literature, we reduced the residual uncertainty of the
short-distance charm-quark contribution to the weak Hamiltonian by
more than an order of magnitude.  The perturbative uncertainty is now
dominated by the missing NNLO corrections to the top-quark
contribution, as well as partially known electroweak corrections at
the percent level (see Refs.~\cite{Gambino:1998rt, Brod:2008ss,
  Brod:2010hi}). The calculation of these corrections~\cite{BGS} has
the potential to bring the perturbative uncertainty of $\epsilon_K$
down to the percent level, motivating a renewed effort to compute
long-distance effects using lattice QCD.

By contrast, the real part of the $|\Delta\rS\!=\!2|$ Hamiltonian is
dominated by up- and charm-quark contributions, and their convergence
is not improved.  Hence, the calculation of these contributions is a
genuine task for lattice QCD, to which a significant effort is
devoted~\cite{Christ:2012se,Bai:2014cva, Blum:2015ywa}.  However, our
results have the potential to supply useful cross checks for part of
these calculations: By performing the matching to the hadronic matrix
elements for $\epsilon_K$ above the charm-quark threshold we can
obtain a prediction of these matrix elements that can be directly
compared to a future lattice calculation. This could shed additional
light onto the lattice calculation of the kaon mass difference.


\begin{acknowledgments}
JB acknowledges support in part by DOE grant DE-SC0020047.
MG is supported in part by the UK STFC under Consolidated Grant ST/L000431/1
and also acknowledges support from COST Action CA16201 PARTICLEFACE.
\end{acknowledgments}

\bibliography{references.bib}

\end{document}